\documentclass{article}

 \usepackage{graphicx}
 \usepackage{float}  
 
\usepackage[creativeai, final]{neurips_2025}

\usepackage[utf8]{inputenc} 
\usepackage[T1]{fontenc}    
\usepackage{hyperref}       
\usepackage{url}            
\usepackage{booktabs}       
\usepackage{amsfonts}       
\usepackage{nicefrac}       
\usepackage{microtype}      
\usepackage{xcolor}         

\PassOptionsToPackage{numbers, compress}{natbib}

\title{‘Studies for’: A Human-AI Co-Creative Sound Artwork Using a Real-time Multi-channel Sound Generation Model}

\author{%
  Chihiro Nagashima , Akira Takahashi , Zhi Zhong , Shusuke Takahashi , Yuki Mitsufuji\\
  Sony Group Corporation, Tokyo, Japan\\
}

\begin{document}

\maketitle

\begin{figure}[H]
    \centering
    \includegraphics[width=0.75\linewidth]{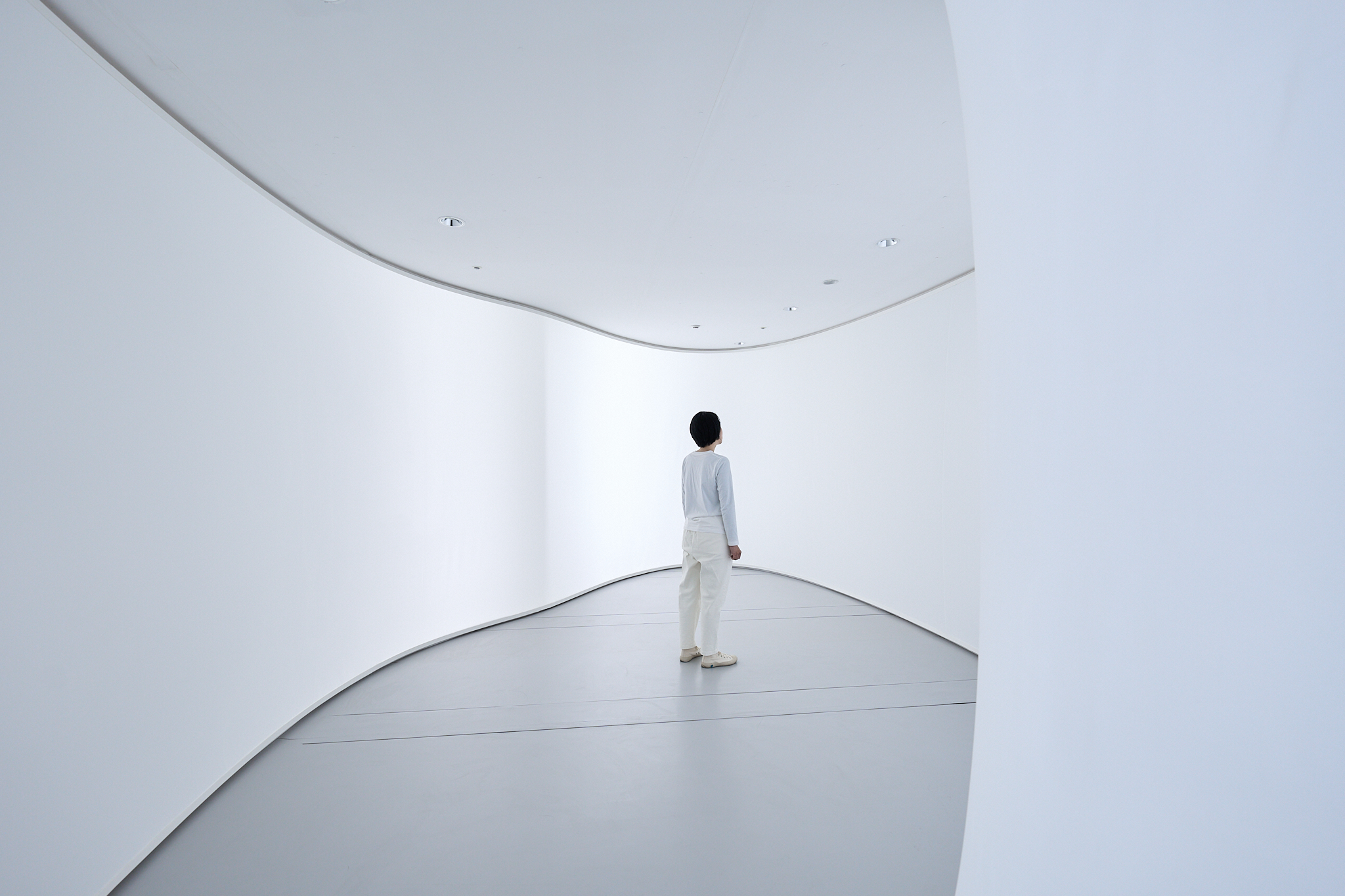}
    \caption{\emph{Studies for}, a collaborative generative sound installation created with sound artist Evala, exhibited at the NTT InterCommunication Center [ICC] in Tokyo from December 14, 2024, to March 9, 2025 (\cite{evala2024studiesfor}). The installation space was enveloped in white fabric, with eight-channel speakers placed behind the fabric. The audience experienced the work by walking through the space. Photo by Maruo Ryuichi, courtesy of ICC.
}
    \label{fig:studies_for}
\end{figure}

\begin{abstract}
  This paper explores the integration of AI technologies into the artistic workflow through the creation of \emph{Studies for}, a generative sound installation developed in collaboration with sound artist Evala (\url{https://www.ntticc.or.jp/en/archive/works/studies-for/}). The installation employs SpecMaskGIT, a lightweight yet high-quality sound generation AI model, to generate and playback eight-channel sound in real-time, creating an immersive auditory experience over the course of a three-month exhibition. The work is grounded in the concept of a "new form of archive," which aims to preserve the artistic style of an artist while expanding beyond artists' past artworks by continued generation of new sound elements. This speculative approach to archival preservation is facilitated by training the AI model on a dataset consisting of over 200 hours of Evala’s past sound artworks.

By addressing key requirements in the co-creation of art using AI, this study highlights the value of the following aspects: (1) the necessity of integrating artist feedback, (2) datasets derived from an artist's past works, and (3) ensuring the inclusion of unexpected, novel outputs. In \emph{Studies for}, the model was designed to reflect the artist's artistic identity while generating new, previously unheard sounds, making it a fitting realization of the concept of "a new form of archive." We propose a Human-AI co-creation framework for effectively incorporating sound generation AI models into the sound art creation process and suggest new possibilities for creating and archiving sound art that extend an artist's work beyond their physical existence. Demo page: \url{https://sony.github.io/studies-for/}

\end{abstract}

\section{Introduction}
Recent advancements in sound generation AI technologies (\cite{kreuk2022audiogen, huang2023makeanaudio2, comunita2024specmaskgit, saito2024soundctm, evans2025sa-open}) have made it possible to generate high-resolution sound. Both AI technologies and AI-driven art creation initially advanced in the realm of visual arts, particularly with image generation. However, with the development of sound generation AI technologies, it is now valuable to introduce these advancements into the domain of sound art and explore the creativity inherent in sound generation AI technology.

This paper presents a case study of incorporating sound generation AI technology into the sound art domain, through the collaborative creation of the sound art installation \emph{Studies for} (\cite{evala2024studiesfor})with sound artist Evala. \emph{Studies for} is a sound art piece that uses SpecMaskGIT (\cite{comunita2024specmaskgit}), a lightweight yet high-quality sound generation AI model, to generate and playback sound in real-time across eight channels simultaneously throughout the three-month exhibition period. Figure 1 shows a photograph of the actual exhibition venue.

The concept behind this work is "a new form of archive." In sound art and media art, due to their spatial and performative nature, archiving poses a significant challenge because these works often rely on specific environments and moments for their full experience, making it difficult to capture their essence in traditional archival formats (\cite{bulut2006soundart,farbowitz2017mediaart}). Evala, who creates many site-specific sound spatialization works(\cite{evala2024profile}), was concerned that much of his work might never be reproduced after his death. We, as his collaborators, saw the potential of this partnership as an opportunity to explore the possibility of using AI models as a “new form of archive.”

In the context of the arts utilizing AI technology, referencing works such as Sofian Audry's Art in the Age of Machine Learning (\cite{audry2021aiart}), the key requirements when artists work with AI can be classified into two elements:
\begin{itemize}
\item Ensuring that the artist’s unique identity is preserved in the output

\item The inclusion of unexpected surprises in the output that the artist did not anticipate
\end{itemize}
By addressing these requirements, the SpecMaskGIT model trained on Evala’s past works and incorporating his feedback, not only reflected his artistic sensibility in the output but also continued to generate previously unheard sounds.

\section{Related Works}
\label{ssec:related_works}
The challenges of archiving site-specific sound art, particularly works that rely on sound spatialization, are significant due to the strong ties between spatial and performative context, which makes them difficult to preserve in conventional archival formats. Zeynep Bulut, in The Problem of Archiving Soundworks (\cite{bulut2006soundart}), highlights how the ephemeral and temporal nature of sound art complicates the process of archiving, especially for works involving spatialization and performance. Similarly, Jonathan Farbowitz’s research on Archiving Computer-based Artworks (\cite{farbowitz2017mediaart}) examines the risks posed by technological shifts to the sustainability of computer-based artworks.
Given this background, the challenges of archiving Evala’s sound artworks become clear from both sound art and media art perspectives. In response to these challenges, the collaborative work \emph{Studies for} was created with the concept of "a new form of archive". This work seeks to offer an innovative solution to the difficulties surrounding the preservation of spatial and site-specific sound art. The concept is discussed in more detail in Section \ref{ssec:archival_concept}.

One approach to ensuring the preservation of artist’s unique identity in the model output is for artists to evaluate the quality of the generated results and iteratively refine the model by incorporating their feedback until a satisfactory outcome is reached. Prior work by \cite{fiebrink2011human} shows that such generation–evaluation–feedback cycles are essential for maintaining creative flow, and that reducing the time required for each iteration helps the model adapt more effectively to the user’s intentions. In our collaboration, we minimized the model size while maintaining high output quality, enabling the artist to repeat this process with minimal delay.
A notable precedent is Fiebrink et al.’s \emph{Wekinator} (\cite{fiebrink2009meta}), an interactive supervised learning tool that supports iterative refinement based on user feedback, though it does not generate audio signals. To our knowledge, no prior work applies sound generation AI models in the way described here. In this context, the SpecMaskGIT model used in Studies for demonstrates how a lightweight yet high-quality Text-to-audio (T2A) model can support such iterative cycles, allowing the artist’s style to be progressively embedded into the generated sound (see Section \ref{ssec:model_architecture}).

Another approach for the identity preservation is for artists to construct original models on the basis of datasets they have curated or collected themselves. This approach allows the model to internalize and reproduce the distinctive features of an artist's style, ensuring that the generated output remains faithful to the artist’s vision while allowing the artist to actively shape the dataset for training(\cite{audry2021aiart}).
Artists such as Shantell Martin and Sarah Schmettmann in Mind the Machine (\cite{martin2017mind}), Songwen Chung in Drawing Operations (\cite{chung2017drawing}), and Anna Ridler in Myriad (Tulips) (\cite{ridler2018myriad}) have used AI models trained on datasets they personally created. By training the models on their original datasets, these artists ensure that their unique artistic practices are reflected in the AI’s outputs.
In our case of \emph{Studies for}, the AI model was exclusively trained on a sound dataset provided by Evala, consisting of over 200 hours of his past works. This focused training ensures that the model retains and reflects his unique sound characteristics, preserving his artistic identity. Sections \ref{ssec:dataset} and \ref{ssec:optimization} provide further details on this methodology.

When artists incorporate AI technology into their creative process, they often expect the technology to produce outputs that exceed their own imaginative capacities. This is reflected in works such as Exquisite Corpus by Songwen Chung (\cite{chung2016exquisite}), Spring Spyre by Laetitia Sonami (\cite{sonami2011spring}), Listener by Suzanne Kite (\cite{kite2018listener}), and Co(AI)xistence by Justine Emard (\cite{emard2017coaixistence}). These artists explore the AI-human interaction within their work, examining how AI models respond to human behavior and how humans, in turn, react to the outputs of the AI. The AI’s ability to produce surprising and sometimes unpredictable outputs often leads to expanding the artist’s creative boundaries.
In \emph{Studies for}, by combining text prompts (the titles of Evala’s past works) with audio inputs (the signature sound that opens all of his sound artworks), a conditioning structure was implemented to ensure that the generated output did not resemble a mere collage of his past works. This structure enables the model to generate sounds that reflect Evala's artistic style while introducing new, previously unheard elements. This methodology is explained in detail in Section \ref{ssec:semantic}.

\section{Artwork Description}
\subsection{Archival Concept}
\label{ssec:archival_concept}
In Evala’s sound installations, which incorporate site-specific sound spatialization, the challenges related to archiving works, as discussed in Section \ref{ssec:related_works}, are particularly pronounced. Evala has expressed concern that most of his works might never be reproduced again after his death. In response to this, \emph{Studies for}, created in collaboration with Evala, represents a speculative attempt to address these archival challenges by training an AI model using an artist’s past works. This approach aims to preserve the artist’s style while continuing to generate the sounds of their work even after their death, proposing a new form of archive.

\subsection{Spatial Composition}
\emph{Studies for} incorporated a spatial design that embodied its core concept. The installation space was enveloped in a white, curved fabric structure, symbolizing both the beginning of life — a womb where sound is born — and the continuation — the afterlife — a realm where sound persists even after the artist is gone. The actual exhibition space is shown in Figure \ref{fig:studies_for}. This duality created an intimate and abstract auditory environment, immersing the audience in a space between the contrast of life and death. Behind the fabric, eight speakers were positioned around the room, allowing the audience to walk freely within the space and experience the generative sound from varying spatial perspectives.

 During the three-month exhibition, the model generated sound across eight channels in real-time simultaneously. A distinctive feature of Evala's artistic practice is the use of multiple speakers to create spatial and site-specific works. Similarly, this installation utilized eight channels to construct an immersive environment where the audience could experience generative sound, unique to the specific time and place, while walking through the space.

\section{Technical Details}
\subsection{Model Architecture}
\label{ssec:model_architecture}
In this artwork, we adopt SpecMaskGIT (\cite{comunita2024specmaskgit}), a lightweight T2A model, as it can directly generate full-band audio signals from either text or audio input. A comparison of model size and performance with other models is shown in Figure \ref{fig:AudioGenerationModel} of the Appendix. Thanks to the fast generation speed, this model allows artists to evaluate the quality of the output and quickly incorporate their feedback into the model’s learning process. Compared to previous T2A models, which often require a significant amount of time for the generation process, this feature enables the artist to be more deeply involved in the model’s development. As a result, the model is able to generate high-quality outputs that better reflect the artist’s input. This makes SpecMaskGIT a good choice for this project, as it integrates the artist’s feedback more effectively into the creative process, resulting in high-quality, artistically meaningful outputs.
 
The overall system comprises two main components: a spectrogram encoder and decoder (SpecVQGAN) that compresses input audio into latent tokens, and a masked generative Transformer (SpecMaskGIT) that reconstructs Mel-spectrogram token sequences from masked token sequences. Details of the training processes for both SpecVQGAN and SpecMaskGIT are illustrated in Figures \ref{fig:SpecVQGAN} and \ref{fig:SpecMaskGIT} in the Appendix (\cite{comunita2024specmaskgit}). 

\subsection{Dataset and Artistic Conditioning}
\label{ssec:dataset}
In this artwork, the model was trained exclusively on audio data extracted from Evala’s past sound artworks, amounting to approximately 200 hours of material. This artist-specific dataset not only aligned with the conceptual intent of creating a “new form of archive,” but also enabled the generative model to internalize and reproduce sound characteristics unique to Evala’s creative style. Furthermore, the interviews with the artist revealed that a minimum sampling frequency of 48kHz was preferred to faithfully deliver the artistic intent into the generated outputs. To achieve this minimum resolution, a high-quality dataset needed to be used for training. As a result, the audio data extracted from Evala’s past works, which was used for training the model, was resampled to 48kHz before being processed.

\subsection{Optimization of Model Complexity and Performance}
\label{ssec:optimization}
As explained in Section \ref{ssec:related_works}, \emph{Studies for} embodies the concept of "a new form of archive" and reflects Evala’s spatially acoustic and site-specific artistic style. To generate sound that can only be experienced by the audience at a specific time and place, real-time eight-channel simultaneous generation was expected instead of pre-generate audio data then playing back in the venue. As mentioned in the previous subsection, the output needed a minimum sampling frequency of 48kHz. Therefore, to achieve fast inference while maintaining high-resolution output, optimizations were made to the SpecMaskGIT model.

In Section \ref{ssec:system}, we provide a detailed description of the equipment setup used for this project. To meet the requirements for real-time generation speed, it was necessary to ensure that the system could operate without delay. For effective spatial acoustic representation, the minimum number of speakers required was determined to be 8 channels, with speakers positioned to surround the listener's left, right, front, back, above, and below, in accordance with the artist's decision. This eight-channel configuration allowed for the intended spatial expression while maintaining the necessary performance for real-time sound generation.

To reduce model complexity while maintaining the desired quality, we optimized both SpecVQGAN and SpecMaskGIT. In the SpecVQGAN training process, the wav2mel mapping was optimized for 48kHz audio, enabling efficient yet high-fidelity spectrogram encoding. Additionally, for real-time sound generation, HiFi-GAN (\cite{kong2020hifigan}), the vocoder previously used in SpecMaskGIT paper (\cite{comunita2024specmaskgit}), was a bottleneck in terms of generation time. Therefore, we replaced it with Vocos (\cite{siuzdak2023vocos}), a vocoder capable of faster processing. In the SpecMaskGIT training process, the number of Transformer blocks was reduced from 24 to 12, decreasing the parameter count to 89.27 million. The model details between the version of SpecMaskGIT trained on the open dataset AudioSet (\cite{gemmeke2017audioset}) and the version trained for \emph{Studies for} are compared in Table \ref{optimization} in the Appendix.

\subsection{Continuous Sound Generation and Semantic Conditioning}
\label{ssec:semantic}
As mentioned in Section 2, the artist sought unexpected outcomes from the model. Initially, the model was trained exclusively on Evala’s past works, which occasionally generated outputs resembling collages of his previous pieces. While these outputs were technically coherent, the artist found them unsatisfying: rather than merely recombining his past works, he hoped to hear sounds that felt new—unheard, yet still true to his style.

In response to this challenge, we modified the model’s conditioning structure. Initially, the model was conditioned only on a sound query of Evala’s signature sound that opens all of his sound artworks. Later, we incorporated textual prompts (the titles of eight of his past works) alongside the audio input using CLAP (Contrastive Language-Audio Pre-training) (\cite{laion2023clap}). As a result, each of the eight output channels is semantically guided by a corresponding title, and sound is continuously generated through an abstract interpretation of the compositional elements of Evala’s past works specifically, the titles and signature sound.

To implement this dual conditioning, we extended the standard Classifier-Free Guidance (CFG) formulation, which is commonly used to balance diversity and alignment in T2A synthesis (\cite{ho2022classifier,comunita2024specmaskgit}). Specifically, we adapted the CFG framework to incorporate both text and audio conditions, following the approach introduced in SpecMaskFoley (\cite{zhong2025specmaskfoley}). The modified CFG objective is defined as:
\begin{equation}
\label{eq:cfg}
    \ell = \ell_\mathrm{uncond} + t[(\ell_\mathrm{text} - \ell_\mathrm{uncond}) + (\ell_\mathrm{audio} - \ell_\mathrm{uncond})],
\end{equation} 
Where $\ell_\mathrm{uncond}$ denotes the logits gained from the audio backbone without CLAP conditioning, $\ell_\mathrm{text}$ denotes the logits obtained from conditioning SpecMaskGIT with CLAP text features, $\ell_\mathrm{audio}$ denotes the logits earned by conditioning SpecMaskGIT with CLAP audio features, and $t$ denotes the CFG scale.

Additionally, to enable sound to be generate uninterruptedly and continuously throughout the three-month exhibition, we adopted an audio outpainting strategy. The model generates 10-second segments in real time, where the latter 5 seconds of each segment are overlapped with the first 5 seconds of the subsequent generation. This overlapping technique ensures temporal coherence and avoids perceptible discontinuities, allowing for a seamless auditory experience.

\begin{figure}[H]
    \centering
    \includegraphics[width=0.96\linewidth]{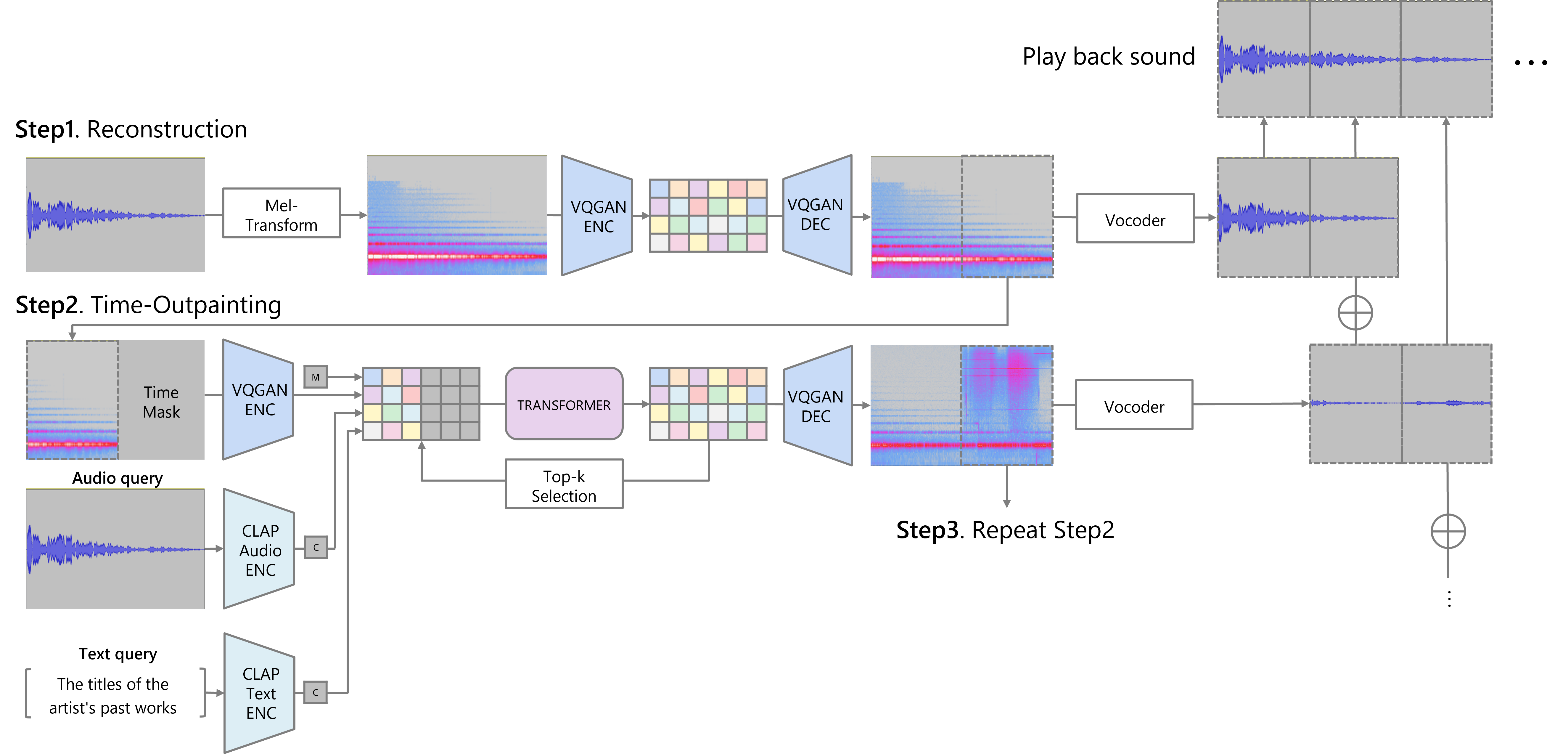}
    \caption{Continuous Generation with Audio and Text Inputs.}
    \label{fig:inference_diagram}
\end{figure}

A diagram of the model structure used in the exhibition is shown in Figure \ref{fig:inference_diagram}.

\subsection{System Implementation and Technical Setup}
\label{ssec:system}
To enable eight-channel audio to be generated and played back in real time at 48kHz, the generative model was deployed a Linux Ubuntu-based workstations, equipped with two NVIDIA RTX 4080 GPUs. The generated audio signals were routed through an RME Babyface interface and a macOS-based machine for sound tuning, including volume balancing and equalization. After processing, the signals were transmitted via an RME Fireface 802 interface to a Crown CT8150 amplifier, which powered the speaker system.

The exhibition space, enclosed in curved white fabric, featured a spatial speaker array with seven Gallo A’DIVA speakers and one Genelec subwoofer positioned behind the fabric, surrounding the audience. Throughout the three-month exhibition, this setup continuously delivered generative eight-channel audio in real time, creating an immersive and ever-evolving auditory environment.A diagram of the equipment used in the actual exhibition is included as Figure \ref{fig:equipment} in the Appendix.

\section{Exhibition Details}
The artwork \emph{Studies for} was exhibited as part of sound artist Evala’s large-scale solo exhibition “Evala: Emerging Site / Disappearing Sight,” held at ICC in Tokyo from December 14, 2024, to March 9, 2025. The exhibition attracted more than 20,000 visitors. While many of Evala’s sound artworks are often presented in relatively dark rooms to make the audience focus on the auditory experience, \emph{Studies for}, in contrast, was displayed in a bright white space (in Figure \ref{fig:studies_for}), which expressed its concept through the environment.

\emph{Studies for} was encountered at the end of the exhibition route, following other works by Evala. The audience expressed that this layout effectively conveyed the idea that the sounds of the piece were generated by a model that had learned from Evala’s previous works. They found this arrangement to be a clear and effective way to understand the relationship between the model and the artist’s sound.

\section{Conclusion}
This study explored the co-creative integration of AI into the artistic workflow of \emph{Studies for}, a generative sound installation created in collaboration with sound artist Evala. To reflect the artist’s intent in the AI model during the artistic production process, it is essential to:
\begin{itemize}
\item Use a lightweight model that allows for quick iterations of trial and error, ensuring that the artist’s identity is preserved in the output while also enabling effective integration of feedback to refine the model and maintain quality. Additionally, using the artist’s past works as the dataset helps reflect the artist’s style in the output.

\item Optimize the system to allow multiple queries as input. This approach introduces novelty and fresh elements that can lead to unexpected surprises, as desired by the artist. This structure, which increases the level of abstraction in the output, was identified as necessary in certain cases.
\end{itemize}
The use of a model like SpecMaskGIT, which can generate audio signals at high speed, enabled sound to be generated and played back in real time across eight channels simultaneously, consistently over a span of three months without any interruptions. This approach is considered an unprecedented example in the field.

In media art, especially in sound art, the work itself is often intangible, and the challenge of archiving such works remains unresolved. However, by training an AI model on an artist’s past works, \emph{Studies for} suggests a speculative new form of archive that not only preserves and references the artist’s past creations but also allows for them to be expanded even after the artist’s death. In this sense, \emph{Studies for} becomes a meaningful work in the context of posthumanism, offering a forward-thinking approach to using AI technology in the artistic domain, and presenting new possibilities for artistic creation beyond the artist’s lifetime.

\clearpage

\bibliographystyle{unsrtnat}
\bibliography{refs25}

\appendix

\section{Technical Appendices and Supplementary Material}
\subsection*{A.1 Comparison of Model Size and Audio Synthesis Performance}
\begin{figure}[H]
    \centering
    \includegraphics[width=1\linewidth]{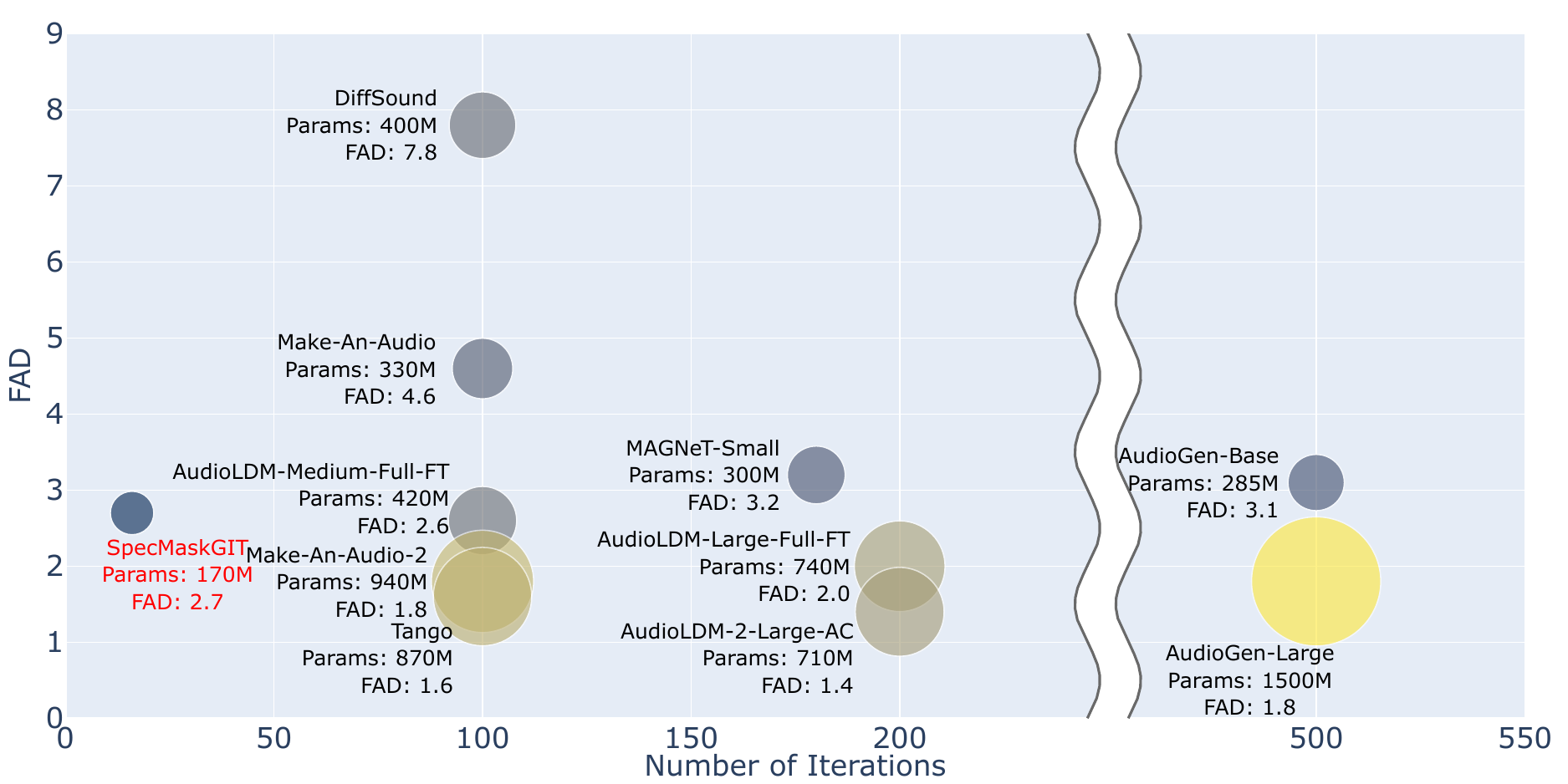}
    \caption{Audio synthesis performance and number of synthesis iterations of different methods. The size of circle represents the model size. SpecMaskGIT achieves decent quality with only 16 iterations and a small model size.}
    \label{fig:AudioGenerationModel}
\end{figure}

\subsection*{A.2 SpecVQGAN Training}

\begin{figure}[H]
    \centering
    \includegraphics[width=1\linewidth]{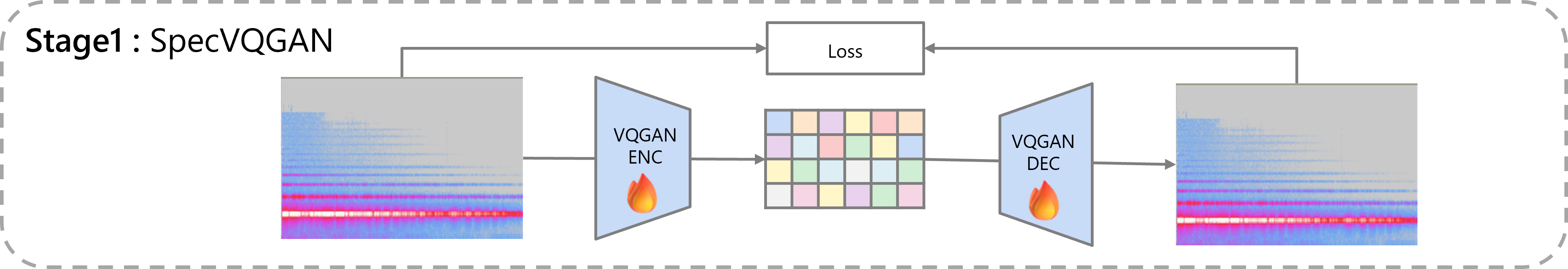}
    \caption{ SpecVQGAN, which encodes non-overlapping
16-by-16 time-mel patches into discrete tokens, and decodes the discrete tokens back to Mel-spectrogram}
    \label{fig:SpecVQGAN}
\end{figure}

\subsection*{A.3 SpecMaskGIT Training}

\begin{figure}[H]
    \centering
    \includegraphics[width=1\linewidth]{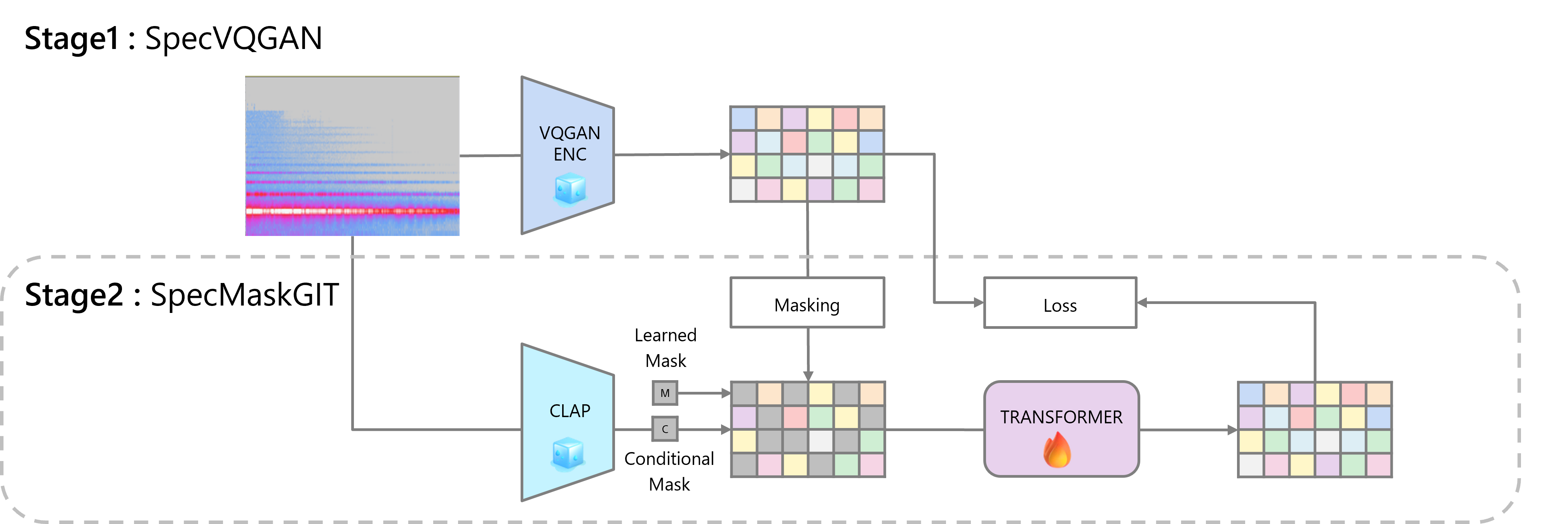}
    \caption{Self-supervised training of SpecMaskGIT. The
Transformer is trained to reconstruct SpecVQGAN token
sequences that are randomly masked with variable masking ratios, conditioned by a semantic embeddding from
the CLAP encoder. “M” denotes the learned mask token,
while “C” denotes the proposed conditional mask.
}
    \label{fig:SpecMaskGIT}
\end{figure}

\begin{table}[H]
  \caption{A comparison of the model details}
  \label{optimization}
  \centering
  \begin{tabular}{c|cc}
    \toprule
         & SpecMaskGIT (\cite{comunita2024specmaskgit})     & Studies for \\
    \midrule
    Dataset & AudioSet 5000 hours & Artist’s original dataset 200 hours \\
    Sampling rate & 22kHz & 48kHz\\
    SpecVQGAN & params: 72M & params: 72M  \\
     & 80 Mel bin x 848 frames & 256 Mel bin x 960 frames\\ 
     & 265 tokens (5 x 53) & 960 tokens (16 x 60) \\
     & 820x compression & 480x compression \\
    SpecMaskGIT & params : 172M  & params : 89M \\
     & 24 Transformer blocks & 12 Transformer blocks \\
     & 768 dim 8 heads attention & 768 dim 12 heads attention \\
     CLAP & 630k-audioset-best.pt & music\_audioset\_epoch\_15\_esc\_90.14.pt \\
     Vocoder & HiFi-GAN & Vocos \\
    \bottomrule
  \end{tabular}
\end{table}

\subsection*{A.4 Equipment Configuration}

\begin{figure}[H]
  \centering
  \includegraphics[width=0.9\linewidth]{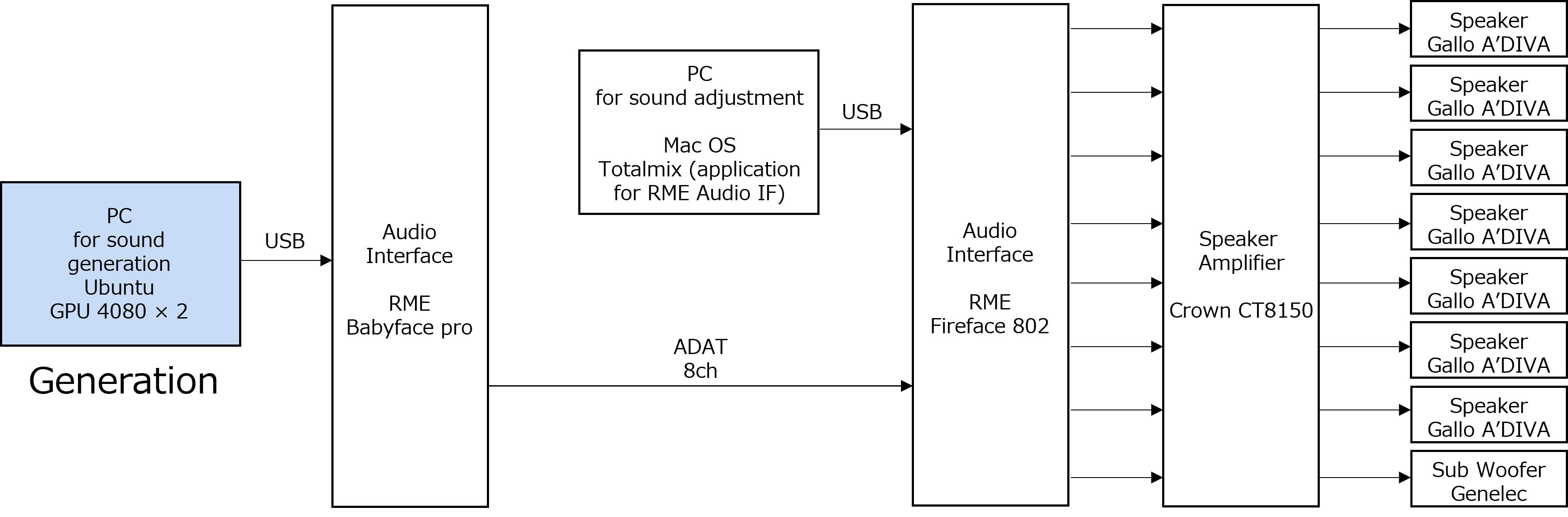}
  \caption{System architecture for real-time eight-channel sound generation using dual RTX4080 GPUs, audio interface (RME Babyface and Fireface802), and spatial speaker setup.}
  \label{fig:equipment}
\end{figure}

\end{document}